\documentclass[a4paper12pt]{article}
\title{On  separable states   for   composite  
  systems of distinguishable fermions}
\date{}
\author{Hajime Moriya 
\thanks{Mailing address: 
e-mail:hmoriya@math.sci.hokudai.ac.jp}}
\usepackage{latexsym}
\usepackage{amssymb} 
\usepackage{amsmath} %
\usepackage{amsthm} %
\usepackage{amscd} %
\newcommand{\R} {{\mathbb{R}}}%
\newcommand{\CC} {{\mathbb{C}}}%
\newcommand{\NN}{{\mathbb{N}}}%
\newcommand{\qedb}{\hbox{\rule[-2pt]{3pt}{6pt}}}%
\newcommand{\I}{{\mathrm{I}}}%
\newcommand{\J}{{\mathrm{J}}}%
\newcommand{\Af}{A_{1}}%
\newcommand{\As}{A_{2}}%
\newcommand{\Bf}{B_{1}}%
\newcommand{\Bs}{B_{2}}%
\newcommand{\Afp}{A_{1+} }%
\newcommand{\Afm}{A_{1-} }%
\newcommand{\Afpm}{A_{1\pm} }%
\newcommand{\Kf}{K_{1}}%
\newcommand{\Ks}{K_{2}}%
\newcommand{\Kfd}{\Kf^{\dagger}}%
\newcommand{\Ksd}{\Ks^{\dagger}}%
\newcommand{\Asp}{A_{2+} }%
\newcommand{\Asm}{A_{2-} }%
\newcommand{\Aspm}{A_{2\pm} }%
\newcommand{\Al}{{\cal{A}}}%
\newcommand{\Alf}{\Al_{1}}%
\newcommand{\Als}{\Al_{2}}%
\newcommand{\AlI}{{\Al}({\I})}%
\newcommand{\AlJ}{{\Al}({\J})}%
\newcommand{\AlJp}{{\cal A}(\J)_{+}}%
\newcommand{\AlJm}{{\cal A}(\J)_{-}}%
\newcommand{\AlIp}{{\cal A}(\I)_{+}}%
\newcommand{\AlIm}{{\cal A}(\I)_{-}}%
\newcommand{\AlIpm}{{\cal A}(\I)_{\pm}}%
\newcommand{\AlJpm}{{\cal A}(\J)_{\pm}}%
\newcommand{\AlJprime}{\AlJ^{\prime}}%
\newcommand{\AlIprime}{\AlI^{\prime}}%
\newcommand{\IuJ}{{\I}\cup {\J}}%
\newcommand{\AlIuJ}{{\Al}(\IuJ)}%
\newcommand{\vp}{\varphi}
\newcommand{\vpl}{\vp_{\lambda}}
\newcommand{\lam}{\lambda}%
\newcommand{\gammat}{\gamma_{\theta}}
\newcommand{\lami}{\lambda_i}%
\newcommand{\ik}{{i(k)}}%
\newcommand{\ij}{{i(j)}}%
\newcommand{\lik}{l_{i(k)}}%
\newcommand{\lij}{l_{i(j)}}%
\newcommand{\ome}{\omega}
\newcommand{\omef}{\ome_{1}}
\newcommand{\omes}{\ome_{2}}
\newcommand{\omei}{\ome_{i}}
\newcommand{\omeih}{{\widehat{\ome}}_{i}}
\newcommand{\ET}{E^{\Theta}}
\newcommand{\ETome}{\ET_{\AlIuJ}(\ome,\AlI,\AlJ)}
\newcommand{\Eomek}{E^{k}_{\AlIuJ}(\ome,\AlI,\AlJ)}
\newcommand{\Eomehalf}{E^{1/2}_{\AlIuJ}(\ome,\AlI,\AlJ)}
\newcommand{\Eomeavr}{E^{{\rm{avr.}}}_{\AlIuJ}(\ome,\AlI,\AlJ)}
\newcommand{\Evarrhoavr}{E^{{\rm{avr.}}}_{\AlIuJ}(\varrho,\AlI,\AlJ)}
\newcommand{\EomeI}{E_{\AlIuJ}(\ome, \AlI)}
\newcommand{\EomeJ}{E_{\AlIuJ}(\ome, \AlJ)}
\newcommand{\EvarrhoI}{E_{\AlIuJ}(\varrho, \AlJ)}
\newcommand{\EvarrhoJ}{E_{\AlIuJ}(\varrho, \AlI)}
\newcommand{\ETvpl}{\ET_{\AlIuJ}(\vpl,\AlI,\AlJ)}
\newcommand{\EvplI}{E_{\AlIuJ}(\vpl, \AlJ)}
\newcommand{\EvplJ}{E_{\AlIuJ}(\vpl, \AlI)}
\newcommand{\Evplavr}{E^{{\rm{avr.}}}_{\AlIuJ}(\vpl,\AlI,\AlJ)}
\newcommand{\omeie}{\ome_{i}^{\text{e}}}
\newcommand{\omefi}{\ome_{1,i}}
\newcommand{\omesi}{\ome_{2,i}}
\newcommand{\omefij}{\ome_{1, i(j)}}
\newcommand{\omesik}{\ome_{2, i(k)}}
\newcommand{\omefih}{\widehat{\ome}_{1,i}}
\newcommand{\omesih}{\widehat{\ome}_{2,i}}
\newcommand{\psih}{\widehat{\psi}}
\newcommand{\omeh}{\widehat{\ome}}
\newcommand{\vi}{v_{i}}%
\newcommand{\vI}{v_{\I}}%
\newcommand{\aicr}{a_i^{\dagger}}%
\newcommand{\ai}{a_i}%
\newcommand{\ajcr}{a_j^{\dagger}}%
\newcommand{\aj}{a_j}%
\newcommand{\afcr}{a_1^{\ast}}%
\newcommand{\af}{a_1}%
\newcommand{\ascr}{a_2^{\ast}}%
\newcommand{\as}{a_{2}}%
\newcommand{\Pl}{P(\lambda)}
\newcommand{\identitybf}{{\mathbf{1} } }
\newcommand{\idbf}{ {\mbox{id}} }
\newcommand{\proofend}{{\hfill \qedb}}
\newcommand{\nonum}{\nonumber}%
\newcommand{\Tr}{\mathbf{Tr}}%
\newcommand{\ket}[1]{|#1 \rangle}
\newcommand{\ketA}[1]{|#1 \rangle_{A}}
\newcommand{\ketB}[1]{|#1 \rangle_{B}}
\newcommand{\ketAB}[1]{|#1 \rangle_{AB}}
\newcommand{\ketbra}[2]{|  #1 \rangle \langle  #2|}
\newcommand{\ketAbra}[2]{|  #1 \rangle_{A} \langle  #2|}
\newcommand{\ketBbra}[2]{|  #1 \rangle_{B} \langle  #2|}
\newcommand{\ketABbra}[2]{|  #1 \rangle_{AB} \langle  #2|}
\newcommand{\Psip}{\Psi_{+}}
\newcommand{\hopterm}{1/2(\Kfd \Ks-\Kf \Ksd)}
\newcommand{\hopnama}{\Kfd\Ks-\Kf \Ksd}
\newcommand{\hopf}{\Kfd\Ks}
\newcommand{\hops}{\Kf\Ksd}
\begin{document}
\maketitle
\theoremstyle{plain}%
\newtheorem{lem}{Lemma}
\newtheorem{pro}[lem]{Proposition}%
\newtheorem{thm}[lem]{Theorem}
\newtheorem{cor}[lem]{Corollary}%
\ \\
\ \\
\ \\
\ \\
{\Large{
\begin{abstract}
We consider 
separable (i.e. classically correlated) states 
for composite
 systems   of   spinless  fermions that are distinguishable.
 For  a proper formulation  of entanglement formation 
  for such systems,
 the state decompositions 
 to be taken should  respect
the univalence superselection rule.   
Fermion hopping  always  induces  non-separability, while
 states with  bosonic hopping correlation 
may or may not be separable.
If we transform  a given 
   bipartite fermion system  into 
 a tensor product one 
by Jordan-Klein-Wigner transformation,
  any separable state for  the former 
 is also  separable for the  latter. 
There are    U(1) gauge invariant  states
that are non-separable for the former but    
separable for the latter.
\end{abstract}
}}
\ \\
\ \\
{\bf {Key Words:}} CAR systems.
 Classically  correlated (separable) states.
Univalence superselection rule.
%
\newpage
\section{Introduction}
\label{sec:INTRO}
We consider    characterization  of      
separable, i.e.  classically correlated states 
 for    lattice fermion systems where 
 fermion particles  on different sites 
are   distinguishable. 
(For the  case of  indistinguishable fermions 
 that are represented as  anti-symmetric wave 
 functions,  see 
e.g. \cite{INDIS4} 
and its  references.)

Let $\NN$ be  
a lattice of integers    ordered by  inclusion.
The canonical anticommutation relations (CARs)  are 
\begin{eqnarray}
\label{eq:CARs}
\{ \aicr, \aj \}&=&\delta_{i,j}\, \identitybf, \nonum \\
\{ \aicr, \ajcr \}&=&\{ \ai, \aj \}=0,\quad \ \ i, j\in \NN,
\end{eqnarray} 
where 
 $\aicr$ and $\ai$  are   creation and  annihilation
spinless fermion  operators 
on the site $i$, and $\{A, B\}=AB+BA$.
For each  subset  $\I$ of $\NN$, 
 the subsystem  $\AlI$ are 
 generated by all 
$\aicr$ and $\ai$ in   $\I$.

 Let   $\I$ and $\J$ be  
disjoint subsets of $\NN$.
We  are interested in characterization of    
  state correlations 
 between  the pair of subsystems  $\AlI$ and $\AlJ$.
We shall comment on our  motivation.
It is sometimes useful  to convert  the argument 
for  quantum spin models    to that for the corresponding 
fermion lattice models by Jordan-Klein-Wigner transformations
 vice versa in quantum statistical mechanics.
We hope that the comparison of   
tensor product systems and CAR systems 
in terms of state correlations
would   be useful   for some  purpose,
 though do not have any  practical suggestion.
Also this work  is  a sort of continuation 
 of \cite{AMEXT} that studied   the independence 
of states  for CAR systems.

We give notation.
The  even-odd grading transformation is  
given  by 
\begin{eqnarray}
\label{eq:THETA}
\Theta(\aicr)=-\aicr, \quad 
\Theta(\ai)=-\ai.  
\end{eqnarray}
The even and odd parts of $\AlI$ are
\begin{eqnarray*}
\AlIpm 
&\:=& \Bigl\{ A \in \AlI \, \bigl|\, \Theta(A)=\pm A  \Bigr\}.
\end{eqnarray*}
%
We  introduce U(1) gauge 
 transformation:
\begin{eqnarray}
\label{eq:GAUGE}
\gammat(\aicr)=e^{i \theta}\aicr, \quad 
\gammat(\ai)=e^{-i \theta}\ai\  
\end{eqnarray}
for $\theta\in \CC^{1}$.
A  state 
that  is  invariant under $\Theta$
 is  called   even,   and a state that  
  is  invariant under $\gammat$ for 
 any 
$\theta\in \CC^{1}$
is  called   U(1)-gauge invariant.

If  the cardinality $|\I|$ 
is  finite, then  $\AlI$
is isomorphic to 
the  $2^{|\I|}\times 2^{|\I|}$ full matrix algebra.
Let 
\begin{eqnarray}
\label{eq:vI}
 \vI := \prod_{i \in \I}\vi,\quad \vi := \aicr\ai-\ai\aicr. 
\end{eqnarray}
This 
$\vI$ gives   
  an even  self-adjoint unitary operator 
implementing $\Theta$, 
\begin{eqnarray}
\label{eq:adv}
   {\mbox {Ad}}(\vI)(A) =\Theta(A),\quad
 A\in \AlI.
\end{eqnarray}


 The notion of  separable states is unchanged  for 
 CAR  systems:
If a state is  written
 as a convex sum of  product states, 
then it is called  a separable   state \cite{WER89}.    
 It is, however,  important  to note that  
due to
 the   CAR  structure (algebraic non-independence)
there are limitations  on  marginal states 
that can be prepared on disjoint regions \cite{AMEXT} and hence 
 on product states.

According to  
 the univalence superselection rule \cite{WWW},
any realizable state 
is $\Theta$-invariant. Thus    noneven states are  
 out of our physical  interest. 
However,  any  even state   has    
noneven-state decompositions (i.e. state decompositions in which 
 there are noneven component states)
 unless it is  pure.
For
 a natural   formulation of entanglement formation
  for even states of CAR systems,  
the state decompositions 
 should be taken from 
 the  even-state space only, not from the whole state space.  
 Such quantity now called the entanglement
 formation under the univalence superselection rule 
is   zero if and only if the given even state is separable
 (Proposition \ref{pro:evencrit}). 
(Later we provide  another definition
 of entanglement formation for CAR systems
 that works for non-even states as well
 but seems not so natural.)

We compare fermion systems to tensor product systems 
 in terms of state correlations.
For fermion systems
any particle  hopping term  
between disjoint subsystems always induces 
non-separability (Proposition  \ref{pro:nohopping}),
 while for  tensor product systems, states with 
 particle hopping correlation  may or not may 
  be  separable.  
We show that any  separable state  for the CAR 
 pair ($\AlI$,  $\AlJ$) is also  separable 
 for the tensor product pair 
($\AlI$,  $\AlIprime$),
where $\AlIprime$ denotes the commutant 
 of $\AlI$ in $\AlIuJ$ (Proposition \ref{pro:NAR}).
It was already noted   in   
  \cite{CABAN}  that 
the set of all separable states for the CAR
 pair  is strictly smaller than
 that for the tensor product pair.  
We reproduce this result 
by  our model 
independent argument, which seems to have 
 some merits.
First the statement is valid 
 for  the infinite-dimensional case as well.
Second   
it is clarified    that 
fermionic correlation due to 
particle  hopping   is responsible
 for this strict inclusion which 
is    realized  in 
  U(1)-gauge invariant state space  as  
 shown in $\S$ \ref{sec:GAUGE} by examples.

In $\S$ \ref{sec:NONEVEN}
we consider   the general 
case including  noneven states  and
provide  a  criterion of separability 
 (Proposition 
\ref{pro:nonevencrit}).

\section{Separability condition
 for  bipartite fermion systems}
\label{sec:SEP} 
We  give   
 a  definition 
 of separability
 for fermion systems.
Let $\I$ and $\J$ be mutually disjoint subsets of $\NN$, 
and  $\ome$ be  a  (not necessarily even) state on  $\AlIuJ$.
We denote 
the   restriction of $\ome$ to $\AlI$ ($\AlJ$) by $\omef$ ($\omes$).
Conversely,  we are given a pair of states 
$\omef$ on $\AlI$ and $\omes$ on $\AlJ$.
If there exists a state $\ome$ on the total system $\AlIuJ$
 such that its restriction to $\AlI$ is equal 
 to $\omef$ and  that 
 to $\AlJ$ is  $\omes$, 
 then 
  $\ome$ is called a state extension of  $\omef$ and $\omes$.
 If
\begin{eqnarray}
\label{eq:EXTpair}
\ome(\Af \As)=\omef(\Af) \omes(\As)
\end{eqnarray}
for all $\Af \in \AlI$ and 
 $\As  \in \AlJ$, then 
such $\ome$ 
 is unique and 
 called  the product state extension of
 $\omef$ and $\omes$
  denoted  $\omef\circ\omes$.
The product property in  the converse order, 
 namely
\begin{eqnarray}
\label{eq:converse}
\ome(\As\Af)= \omes(\As)\omef(\Af).
\end{eqnarray}
is a consequence of  (\ref{eq:EXTpair}) combined with  CARs
 and  Proposition  \ref{pro:nohopping} below.

We say  that  
a  state $\ome$ of $\AlIuJ$ 
satisfies the separability for the 
 pair of subsystems  $\AlI$ and $\AlJ$,
 or $\ome$ is a separable state  for  $\AlI$ and $\AlJ$,
if there exist  a set of states $\{\omefi\}$ on $\AlI$,
  also that $\{\omesi\}$ on $\AlJ$,
 and  some  positive numbers $\{\lami\}$ such that $\sum_{i}\lami=1$,
 satisfying that
\begin{eqnarray}
\label{eq:sepdef}
\ome(\Af \As)=\sum_{i} \lami  \omefi\circ \omesi(\Af \As)
\end{eqnarray}
for any  $\Af \in \AlI$ and 
  $\As  \in \AlJ$.
This   formula  requires
 the existence of 
the product state 
 $\omefi\circ \omesi$ for each pair of $\omefi$ and $\omesi$.
For  tensor product systems, the existence of  
product  state extension for any given states on disjoint subsystems 
is  automatic, 
  while  for  fermion systems
 it is not always the case \cite{AMEXT}.
\begin{pro}
\label{pro:nohopping}
Let $\I$ and $\J$ be a pair of  disjoint subsets
 and $\ome$ be a state on $\AlIuJ$.
If $\ome$ is  a  separable state 
 for  $\AlI$ and $\AlJ$, then
for any  $\Afm\in \AlIm$ and $\Asm\in \AlJm$,
\begin{eqnarray}
\label{eq:oddodderase}
\ome(\Afm \Asm)=0.
\end{eqnarray}

If $\ome$ is a product state, then 
at least one of 
its restrictions to $\AlI$ and  $\AlJ$
is  even.
\end{pro}
\proof
First we show the second 
statement.  
Let $\ome$ be a product state with 
 its marginal states $\omef$ on $\AlI$
 and $\omes$ on $\AlJ$.
Now suppose that both 
  $\omef$ and $\omes$ are noneven. Hence 
 there are    odd elements $\Afm\in \AlIm$
 and $\Asm\in \AlJm$ such that $\omef(\Afm)\ne0$
 and  $\omes(\Asm)\ne0$.
We are going to derive the contradiction.
By the assumed product property, 
 \begin{eqnarray}
\label{eq:contradict1}
 \omef\circ \omes(\Afm \Asm)=\omef(\Afm) \omes( \Asm)\ne0.
\end{eqnarray}
 Both $\Afm+\Afm^{\dagger}$ and  
$i(\Afm-\Afm^{\dagger})$ are self-adjoint elements 
in $\AlIm$.
Since $\Afm$ can be written as their  
linear combination,
 the expectation value of at least one of them for $\omef$
must be non-zero.
Thus we can  take 
$\Afm=\Afm^{\dagger}\in \AlIm$ such that 
  $\omef(\Afm)\ne 0$ and similarly 
 $\Asm=\Asm^{\dagger}\in \AlJm$ such that 
  $\omes(\Asm)\ne 0$. 

 Now    both $\omef(\Afm)$ and $\omes(\Asm)$
 are non-zero real,  hence 
 $\omef(\Afm) \omes( \Asm)$ is non-zero real.
On the other hand, 
$\Afm\Asm$ is skew self-adjoint as
 \begin{eqnarray*}
(\Afm\Asm)^{\dagger}=\Asm^{\dagger}\Afm^{\dagger}=\Asm\Afm
=-\Afm\Asm.
\end{eqnarray*}
Thus 
$\omef\circ \omes(\Afm\Asm)$ must be purely  imaginary,
which is  a contradiction.

We  assume that $\ome$ is a separable state.
By definition, $\ome$
 has a decomposition 
 into the affine sum of product states:
\begin{eqnarray*}
\ome=\sum_{i} \lami  \omefi\circ \omesi.
\end{eqnarray*}
Suppose that
there exist 
 $\Afm\in \AlIm$ and $\Asm\in \AlJm$ such that
\begin{eqnarray*}
\label{eq:}
\ome(\Afm \Asm)\ne 0.
\end{eqnarray*}  
 Then there exists some product state  $\omefi\circ \omesi$
 in the decomposition such that 
 \begin{eqnarray*}
\label{eq:}
 \omefi\circ \omesi(\Afm \Asm)\ne0.
\end{eqnarray*}
But this is impossible.  Our 
 assertion is now proved.
\proofend
\ \\
\ \\
%

For 
 a given symmetry $G$, 
   there may  exist   $G$-invariant separable states
 which 
have   no   separable decomposition
 that consists of all   $G$-invariant product  states 
 \cite{VERC}, for example, U(1)-symmetry.
The next proposition 
shows the nonexistence of  such separable states
  for  $\Theta$-symmetry.
\begin{pro}
\label{pro:loccsep}
Let $\I$ and $\J$ be a pair of  disjoint subsets and 
$\ome$ be 
an even  state on $\AlIuJ$.
If $\ome$ is a separable state
 for $\AlI$ and $\AlJ$, then
 it has a  separable decomposition 
\begin{eqnarray}
\label{eq:evendec}
\ome=\sum_{i} \lami  \omefi\circ \omesi,
\end{eqnarray}
 such that 
  $\lami>0$, $\sum_{i}\lami=1$, and 
all the marginal states  $\omefi$ on $\AlI$
 and $\omesi$ on $\AlJ$ are even.

If  $\I$ and $\J$ are finite subsets,
  all $\omefi$ and $\omesi$  above can be taken from 
 the set of pure  even  states.
\end{pro}
\proof
Let   $\ome=\sum_{i} \lami  \omei$
 where  $\omei:= \omefi\circ \omesi$,
$\omefi$ and $\omesi$ are some states on $\AlI$ and $\AlJ$.
We shall show  that all 
$\omefi$ and $\omesi$ can be taken  from  even states.

By Proposition \ref{pro:nohopping}
at least one of $\omefi$ and  $\omesi$ should be  even
  for the existence of the
 product state $\omefi\circ \omesi$.
For a given state $\psi$ 
 let $\psih$ denote its  
 $\Theta$-averaged state $\frac{\psi+\psi \Theta}{2}$.
By the evenness of $\ome$, we have the following identity:
\begin{eqnarray*}
\label{eq:omei}
\ome=\omeh=\sum_{i} \lami \omeih.
\end{eqnarray*}
For each $i$, 
$\omeih$ is   an  even 
product state for  $\AlI$ and $\AlJ$
 because $\omeih= \omefih \circ \omesih$.
Replacing $\omefi$ and $\omesi$ by 
$\omefih$ and $\omesih$, we obtain
 a  separable decomposition for $\ome$
 consisting of all even states.

For a finite dimensional CAR system, 
 every  even state can be  decomposed 
 into an  affine sum of pure even states. 
Hence 
if $\I$ is  finite,
we have $\omefi=\sum_{\ij}\lij\omefij$,
 where $\lij>0$, $\sum_{\ij}\lij=1$  and 
 each  $\omefij$ is a   pure even state of $\AlI$.
 Similarly,   $\omesi=\sum_{k}\lik\omesik$,
 where  $\lik>0$, $\sum_{\ik}\lik=1$  and 
 each  $\omesik$ is a    pure even  state of $\AlJ$.
Hence we have an  even-pure-state decomposition 
$\omefi\circ \omesi=\sum_{\ij,\ik}\lij \lik  
\omefij\circ \omesik$  
for each  $i$.
Those for  all  indexes 
induce   a  desired decomposition of $\ome$. 
\proofend
\ \\

For the second statement of this proposition,
 the assumption that $\I$ and $\J$
are finite subsets is necessary since 
 there  is an even state that 
 is pure on $\AlIp$ but non-pure on $\AlI$ 
when    $|\I|$ is infinite
\cite{MANVER}.

\ \\
\noindent{{\it Remark 1:}}
Examples of bosonic U(1)-gauge invariant 
  separable states  
that cannot be prepared locally
 under the U(1)-gauge symmetry  
are given
in  the above mentioned  reference \cite{VERC}.
We now  consider the lattice-fermionic counterpart of 
 Example 1
(eq.4) given there.
Let   $\ket {0}$  and $\ket{1}$
 be  the unit vector denoting  the absence  and  the 
presence
 of one-fermion particle.
 Let  two disjoint
 subsystems under consideration be indicated by  $A$ and $B$.
 Let
\begin{eqnarray}
\label{eq:rhoone}
\rho_1:=  \frac{1}{4}\Big(
\ketAbra{0}{0} \otimes 
\ketBbra{0}{0}+
\ketAbra{1}{1} \otimes 
\ketBbra{1}{1} \Bigr)+
1/2 \ketABbra{\Psip}{\Psip},
\end{eqnarray}
 where $\ketAB{\Psip}:=\frac{1}{\sqrt{2}}
(\ketA{0} \ketB{1}+ \ketA{1} \ketB{0})$.
 Let  $ \ket{a_{1,2}} = \ket{b_{1,2}}:=
\frac{1}{\sqrt{2}}(\ket {0}  \pm \ket{1} )$ and
    $ \ket{a_{3,4}} = \ket{b_{3,4}}:=
\frac{1}{\sqrt{2}}(\ket {0}  \pm i \ket{1} )$,
 where $a$ and $b$ indicate that 
the states are of $A$
 and $B$, respectively, and the subscripts 
$1$ and $2$ correspond  to $+$ and $-$, respectively.  

For the bosonic  case, 
 $\rho_1$ is separable since  it has  
 its separable 
 decomposition: 
$\rho_1=\sum_{k=1}^{4}
\ketbra{a_{k}}{a_{k}} \otimes 
\ketbra{b_{k}}{b_{k}}$.

For the fermionic case,
$\rho_1$  is nonseparable.
Note that  the notation  
$\ketbra {a_{k}}{a_{k}} \otimes 
\ketbra{b_{k}}{b_{k}} $ ($k=1,2,3,4$) 
 makes no sense  (even mathematically), because 
there is   no product state extension
 for   $\ketbra {a_{k}}{a_{k}}$
 on $A$
 and $\ketbra{b_{k}}{b_{k}} $ on $B$
that are both noneven states.
 (In fact  
 there is no state extension at all for them
by  Theorem 1 (2) of \cite{AMEXT}.)
Furthermore,  Proposition  \ref{pro:nohopping} 
claims  the nonexistence of 
 separable decomposition of
$\rho_1$ due to the 
particle hopping correlation by   
$\ketABbra{\Psip}{\Psip}$.
\ \\

Let 
$\AlIprime$  ($\AlJprime$) denote  the commutant 
 algebra of $\AlI$  ($\AlJ$) in $\AlIuJ$.
If the cardinality $|\I|$
 of $\I$ is infinite, $\AlIprime=\AlJp$.
If $|\I|$ is finite, $\AlIprime=\AlJp+\vI\AlJm$ and 
$\AlIuJ= \AlI \otimes \AlIprime$ hold.
As is well known,   the CAR pair $(\AlI, \AlJ)$
 is transformed to the tensor product pair
$(\AlI, \AlIprime)$  
and to $(\AlJ, \AlJprime)$  
by  Jordan-Klein-Wigner transformations. 
We consider    how 
 the properties  of state correlation (separability,  
 entanglement degrees, etc) will remain or change
 by  the replacement of   the CAR pair 
by  the tensor-product ones, and  vice versa. 
The following proposition 
 shows that the  separability  condition  for  
the CAR pair always implies that for the tensor product pair
for even states.
We have noted   in {Remark 1} that 
the converse  of this proposition 
does not hold. 
Later in Proposition 
\ref{pro:GNA}
we will see
 that the evenness assumption is unnecessary.
We   now provide the  simple  proof that 
 makes use of the evenness assumption.
\begin{pro}
\label{pro:NAR}
Let $\I$ and $\J$ be a pair of  disjoint subsets and 
$\ome$ be 
an even  state on $\AlIuJ$.
If it is  separable 
 for  the CAR pair $\AlI$  and   $\AlJ$, then  
 so it is   for the tensor product pair 
 $\AlI$ and    $\AlIprime$.
\end{pro}
\proof
Since   $\ome$ is  an even   separable state,
 it has a separable decomposition in the form 
 of (\ref{eq:evendec}) where each   
 $\omefi$ and $\omesi$  is   even.
By CARs and the evenness of $\omefi$ and $\omesi$, 
we verify that 
 $\omefi\circ \omesi$ is a product state 
with respect to  the tensor product pair 
 $\AlI$ and    $\AlIprime$.
 Hence the separability of $\ome$
 for the pair ($\AlI$,  $\AlIprime$) follows.
\proofend
\ \\

\section{The entanglement formation 
 under the univalence superselection rule} 
\label{sec:ENTAN}
We  introduce a quantity
 that  measures non-separability
 of even  states between $\AlI$ and $\AlJ$ for  
 disjoint finite subsets $\I$ and $\J$. 
The   von Neumann entropy
of the density matrix $D$ is given  by
\begin{eqnarray}
\label{eq:vn}
-\Tr\bigl( D \log D  \bigr),\
\end{eqnarray}
where $\Tr$ denotes   the  trace  which takes the value $1$ 
 on each minimal projection.
The von Neumann entropy of a state 
$\ome$ is given  by (\ref{eq:vn}) for  its density 
 matrix with respect to $\Tr$ 
and is denoted   $S(\ome)$.

For     even state $\ome$ of $\AlIuJ$,
we define
\begin{eqnarray}
\label{eq:ETome}
\ETome&:=&
\inf_{\ome = \sum \lami \omeie }
 \sum_{i} \lami    S(\omeie|_{\AlI}),
\end{eqnarray}
 where the infimum is taken over   all  even-state decompositions 
 of  $\ome$.
   Namely,  each   $\omeie$ 
is an  even state on $\AlIuJ$.
We shall call this quantity 
 {\it{entanglement of formation under the 
 univalence superselection rule}}.
From \cite{VFCAR}
 it follows that 
\begin{eqnarray}
\label{eq:symmetric}
S(\omeie|_{\AlI})=S(\omeie|_{\AlJ})= S(\omeie|_{\AlIp})
= S(\omeie|_{\AlJp}).
\end{eqnarray}
Thus 
the subsystem  in the r.h.s of (\ref{eq:ETome})
can be  any of  $\AlI$, $\AlJ$, $\AlIp$ and $\AlJp$.
We  give  a criterion of the separability
between the CAR pair $\AlI$ and $\AlJ$
 in terms of this degree.
 \begin{pro}
\label{pro:evencrit}
Let  $\I$ and $\J$ be  finite disjoint 
  subsets and
 $\ome$ be an even  state of $\AlIuJ$. 
It is a separable state  
   for  $\AlI$ and $\AlJ$ 
if and only if its  entanglement formation
 under the univalence superselection rule    
$\ETome$ is equal to zero.
\end{pro}
\proof
If an even state $\ome$ satisfies the separability condition,
 then by Proposition 
\ref{pro:loccsep}
there exists a product-state decomposition
\begin{eqnarray}
\label{eq:sepagain}
\ome(\Af \As)=\sum_{i} \lami  \omefi\circ \omesi(\Af \As)
\end{eqnarray}
 such that each  of $\omefi$ and $\omesi$
 is  even and pure.
Thus $\ETome=0$ by definition.
The converse direction is easily verified.
\proofend
%
\section{
Non-separable for the CAR pair 
 but  separable 
 for the tensor product pair} 
\label{sec:GAUGE}
We construct some  U(1)-gauge invariant  states
that are separable for the tensor product pair 
 ($\AlI$,  $\AlIprime$)
but non-separable 
for the CAR pair 
($\AlI$,  $\AlJ$).
 As will be specified below,  their  non-separability is     
purely due to   fermion  hopping terms.

Let $\tau$ be the tracial state  on $\AlIuJ$.
We note the  following product properties of the tracial state: 
\begin{eqnarray}
\label{eq:proIJ}
\tau(\Af\As)&=&\tau(\Af)\tau(\As),\nonum\\
\end{eqnarray}
 for every  $\Af\in\AlI$ and $\As \in \AlJ$,
 and 
\begin{eqnarray}
\label{eq:procommu}
\tau(\Af\Bs)&=&\tau(\Af)\tau(\Bs),\nonum\\
\tau(\Bf\As)&=&\tau(\Bf)\tau(\As),\nonum\\
\end{eqnarray}
 for every $\Af\in \AlI$, 
$\Bs\in \AlIprime$, and every 
 $\Bf \in \AlJprime$,
$\As \in \AlJ$.

Let $\Kf$ and  $\Ks$ be odd elements 
in $\AlIm$ and in $\AlJm$.
Typically 
those are  field operators on 
specified regions.  
Let  $K:=\hopterm$,  which is self-adjoint
 and may  represent    fermion hopping.
Suppose that $\Vert \Kf \Vert\le 1$
  $\Vert \Ks  \Vert\le 1$, 
 then   $\Vert K  \Vert\le 1$.
For $\lam\in \R$, define 
\begin{eqnarray}
\label{eq:Pl}
\Pl := \idbf+ \lam K.
\end{eqnarray}
By definition   $\Pl$ is self-adjoint, and by
\begin{eqnarray*}
\label{eq:bound}
\Vert \lam K \Vert\le |\lam|,
\end{eqnarray*}
 it is a positive operator 
  if $|\lam|\le 1$.
 From   (\ref{eq:proIJ}) and the evenness
of the  tracial state it follows that 
\begin{eqnarray}
\label{eq:tauPl}
\tau(\Pl)&=&\tau(\idbf+ \lam K )=\tau\Bigl(\idbf+ \frac{\lam}{2} 
(\hopnama) \Bigr)\nonum\\
&=&
\tau(\idbf)+\frac{\lam}{2} \bigl( \tau( \hopf)-\tau(\hops)\bigr)\nonum\\ 
&=&\tau(\idbf)+
\frac{\lam}{2} \bigl( \tau(\Kf^{\dagger} )\tau(\Ks)
-\tau(\Kf)\tau(\Ks^{\dagger})
\bigr)  \nonum\\
&=&\tau(\idbf)+\frac{\lam}{2} \cdot  0=1.
\end{eqnarray}
Hence for $\lam\in \R$,  $|\lam|\le 1$, 
$\Pl$ is  a density matrix with respect to the tracial 
 state $\tau$.
 Let us define the state $\vpl$ on $\AlIuJ$ by
\begin{eqnarray}
\label{eq:vpl}
\vpl(A)&:=& \tau(\Pl A),\quad A\in\AlIuJ.
\end{eqnarray}
By definition, 
\begin{eqnarray*}
\Theta(\Pl)=\Pl,
\end{eqnarray*}
hence
$\vpl$ is an even state of $\AlIuJ$.

We  now compute the expectation value
of $\vpl$ for  the  product element 
$\Af\As$ of  $\Af  \in \AlI$  and 
$\As  \in \AlJ$.
We have 
\begin{eqnarray*}
\label{eq:}
\tau\left((\hopf)  \Af\As\right) 
&=&\tau\left(\Kfd(\Ks \Af) \As\right) \nonum\\
&=&\tau\left(\Kfd  ( \Theta(\Af) \Ks) \As\right)=
\tau\left((\Kfd\Theta(\Af))( \Ks \As)\right)  \nonum\\
&=&\tau \left( (\Kfd\Theta(\Af) \right)\tau( \Ks \As)\nonum \\
&=&\tau \circ \Theta\left( \Theta(\Kfd) \Af \right)\tau( \Ks \As)\nonum \\
&=&\tau \left( \Theta(\Kfd) \Af \right)\tau( \Ks \As)\nonum \\
&=&\tau \left((-\Kfd) \Af\right)\tau( \Ks \As)\nonum \\
&=&-\tau (\Kfd \Af)\tau( \Ks \As),
\end{eqnarray*}
and similarly
\begin{eqnarray*}
\label{eq:hoppterms}
\tau\left((\hops)  \Af\As\right) 
&=&-\tau (\Kf \Af)\tau( \Ksd \As),
\end{eqnarray*}
where we have used   CARs, 
(\ref{eq:adv}), (\ref{eq:proIJ}), 
and $\tau=\tau\circ\Theta$ 
 which follows from 
the uniqueness of  the tracial state.
Thus we obtain
\begin{eqnarray*}
\label{eq:}
\vpl(\Af\As)&=& \tau(\Af\As) -\frac{\lam}{2}
\Bigl(
\tau (\Kfd \Af)\tau( \Ks \As)-
\tau (\Kf \Af)\tau( \Ksd \As)
\Bigr).
\end{eqnarray*}
 Since the tracial state is an even product state and 
 $\Kf\in \AlIm$, $\Ks\in \AlJm$,
writing   
$\Af=\Afp+\Afm$, 
$\Afpm \in \AlIpm$,   
$\As=\Asp+\Asm$, 
$\Aspm \in \AlJpm$,
 we obtain 
\begin{eqnarray*}
\label{eq:afas1}
\vpl(\Af\As)=
\tau(\Afp)(\Asp)- 
\frac{\lam}{2}
\left(
\tau(\Kfd\Afm)\tau( \Ks \Asm)-\tau(\Kf\Afm)\tau( \Ksd \Asm)
\right).
\end{eqnarray*}
Similarly we have 
\begin{eqnarray*}
\label{eq:asaf}
&&\vpl(\As\Af)\nonum\\
&=&
\tau(\Af\As) +
\frac{\lam}{2}
\Bigl(
\tau (\Kfd \Af)\tau( \Ks \As)-
\tau (\Kf \Af)\tau( \Ksd \As)
\Bigr) \\
&=&
\tau(\Afp)(\Asp)
+ \frac{\lam}{2}
\left(
\tau(\Kfd\Afm)\tau( \Ks \Asm)-\tau(\Kf\Afm)\tau( \Ksd \Asm)
\right).
\end{eqnarray*}
We summarize the above computations as follows.
\begin{pro}
\label{pro:vpl}
The state $\vpl$ given by   the density 
$\Pl := \idbf+i \lam K$
 with   
$\lam\in \R, \;|\lam|\le 1$, 
$K:=\hopterm$, 
$\Kf
 \in \AlIm$ and $\Ks \in \AlJm$
 such that  $\Vert \Kf \Vert\le 1$
 and  $\Vert \Ks  \Vert\le 1$, 
 has the following correlation functions:
\begin{eqnarray}
\label{eq:lam1}
\vpl(\Af\As)\!\!\!\!\!\!&=&\!\!\!\!\!\!
\tau(\Afp)(\Asp)\!-\! \frac{\lam}{2}\!\!
\left(
\tau(\Kfd\Afm)\tau( \Ks \Asm)
\!\!-\!\! \tau(\Kf\Afm)\tau( \Ksd \Asm)
\right), \nonum\\
\vpl(\As\Af)\!\!\!\!\!\!&=&\!\!\!\!\!\!
\tau(\Afp)(\Asp)\!+ \!\frac{\lam}{2}\!\!
\left(
\tau(\Kfd\Afm)\tau( \Ks \Asm)
\!\!-\!\! \tau(\Kf\Afm)\tau( \Ksd \Asm)
\right).
\end{eqnarray}
\end{pro}
\ \\

Let us  recall a well known criterion of separability
 for tensor product systems in   \cite{HHHSEP}:
 A state is separable
for a bipartite tensor product system
 $\Alf\otimes\Als$
 if and only if it is mapped to a positive 
 element  under $\Lambda \otimes\idbf$ for any 
positive map $\Lambda$ from  
 $\Alf$ to $\Als$. 
By applying this criterion 
 to the density (\ref{eq:Pl}) of $\vpl$, 
we verify that it is  separable  
for  $(\AlI, \AlIprime)$  
and also for  $(\AlJ, \AlJprime)$  
  for any $\lam\in \R, \;|\lam|\le 1$. 
But this is not the case 
for the CAR pair $(\AlI, \AlJ)$. 
Take  one-site subsets
$\I=\{1\}$ and  $\J=\{2\}$, and let
$\Kf=\af$,
$\Ks=\as$ for  computational simplicity.
Then  we have 
\begin{eqnarray}
\label{eq:lam2}
\vpl(\afcr \as)&=&\frac{\lam}{8}\nonum \\
\vpl(\af \ascr)&=&-
\frac{\lam}{8}.
\end{eqnarray}
By Proposition  \ref{pro:nohopping},
$\vpl$ is non-separable between $\AlI$ and $\AlJ$ 
for any non-zero $\lam$.
\section{The general  case 
including  noneven states} 
\label{sec:NONEVEN}

In this section, 
 our  state $\ome$ on $\AlIuJ$
 can be  noneven.
We define the 
following quantity 
 for positive number  $k$, $0\le k \le 1$:   
\begin{eqnarray}
\label{eq:Eomek}
\Eomek :=
\inf_{\ome = \sum \lami \omei }
 \sum_{i} \lami   
 \Bigl(k S(\omei|_{\AlI})+(1-k)(\omei|_{\AlJ}) \Bigr),
\end{eqnarray}
where the infimum is taken over all the state decompositions 
of  $\omega$ in the state space of $\AlIuJ$.
For any pure state $\ome$ of $\AlIuJ$, it reduces to
\begin{eqnarray}
\label{eq:EIJpure}
\Eomek=
kS(\ome|_{\AlI})+(1-k) S(\ome|_{\AlJ}).
\end{eqnarray}

For $k=1, 0$, (\ref{eq:Eomek})
 reduces to the usual definition
 of entanglement formation \cite{BEN} denoted 
$\EomeI$ and $\EomeJ$, respectively.
We note that  $\EomeI$ 
quantifies the non-separability of states 
 for the tensor product pair $(\AlI, \AlIprime)$,
 not for our target $(\AlI, \AlJ)$.

Asymmetry of  entanglement may arise   
for  noneven  states as shown  in  \cite{SOME}.  
For example,
 there is a noneven pure state $\varrho$ on $\AlIuJ$
 such that $\varrho|_{\AlI}$ is a pure state while 
$\varrho|_{\AlJ}$ is a tracial state,
 giving 
\begin{eqnarray}
\label{eq:varrho}
0=S(\varrho|_{\AlI})< S(\varrho|_{\AlJ})=\log 2
\end{eqnarray}
 when  $\I=\{1\}$ and  $\J=\{2\}$. 
Hence   for  quantification
 of state  correlation between  $(\AlI, \AlJ)$ for noneven states,
we have  to take both subsystems  
   on $\I$ and  on $\J$  into account.
Here we 
take   the equal probability $k=1/2$ for simplicity,
and denote    $\Eomehalf$  by $\Eomeavr$ which 
 is called the averaged entanglement formation.
\begin{pro}
\label{pro:nonevencrit}
Let  $\I$ and $\J$ be  finite 
 disjoint subsets and
 $\ome$ be a   state on $\AlIuJ$. 
Then it is a separable state  
   for  $\AlI$ and $\AlJ$ 
if and only if the averaged entanglement formation    
$\Eomeavr$ is equal to zero.
\end{pro}
\proof
If $\ome$ satisfies the separability condition (\ref{eq:sepdef}),
 then there exists the product-state decomposition:
\begin{eqnarray}
\label{eq:sepagain}
\ome(\Af \As)=\sum_{i} \lami  \omefi\circ \omesi(\Af \As).
\end{eqnarray}
 For each index $i$, at least  one of   
$\omefi$ and  $\omesi$
 should be even for the existence of the
 product state $\omefi\circ \omesi$ by
  Proposition \ref{pro:nohopping}. So
let $\omefi$ be even. 
Then it can be decomposed as
$\omefi=\sum_{j}\lij\omefij$,
 where $\lij>0$, $\sum_{j}\lij=1$, and 
 all $\omefij$ can be  taken from  pure even states of $\AlI$.
 (This is always possible when  $\I$ is finite.)
We have a decomposition  of $\omesi$ as 
$\omesi=\sum_{k}\lik\omesik$,
 where  $\lik>0$, $\sum_{k}\lik=1$, and 
 all $\omesik$ are    pure  states of $\AlJ$.
Since each $\omefij$ is  an even state of $\AlI$,
 we are given the (unique) product state
extension $\omefij\circ \omesik$  
for any $\ij$ and $\ik$. 
Repeating  the same  machinery
 for all  $i$, we have 
 a  state decomposition 
of $\ome$ into \{$\omefij\circ \omesik\}$
 where each  $\omefij$ and $\omesik$
 is  a pure state.
 Hence  
\begin{eqnarray*}
S(\omefij|_{\AlI})= S(\omesik|_{\AlJ})=0
\end{eqnarray*}
 for every $\ij$, $\ik$. Thus  this decomposition 
 gives
\begin{eqnarray}
\label{eq:Eavrzero}
\Eomeavr=0. 
\end{eqnarray}
 
Conversely, assume (\ref{eq:Eavrzero}).
By  definition, there exists a 
 state decomposition 
$\ome= \sum_{i} \lami \omei$
such that
\begin{eqnarray}
\label{eq:killall}
S(\omei|_{\AlI})= S(\omei|_{\AlJ})=0,
\end{eqnarray}
 for all $i$.
This implies that $\omei$ 
 has pure state restrictions  on both $\AlI$ and $\AlJ$. 
By Theorem 1 (2) in  \cite{AMEXT},
 at least one of ${\omei}|_{\AlI}$ and  
${\omei}|_{\AlJ}$
 should be even for the existence of their  state extension
 $\omei$ on $\AlIuJ$
 and $\omei$ is uniquely given  as 
${\omei}|_{\AlI}\circ {\omei}|_{\AlJ}$.
Hence   $\ome$ can be written as  the affine 
 sum of the product states $\{\omei\}$ and 
 hence  is a separable state.
\proofend
\ \\

 The  following relationships among the  introduced 
 entanglement formations are obvious.
\begin{lem}
\label{lem:INEQ}
For any   state $\ome$ on $\AlIuJ$,
\begin{eqnarray}
\label{eq:half}
\Eomeavr
\ge 1/2 \EomeI + 1/2 \EomeJ.
\end{eqnarray}
For any  even state $\ome$ on $\AlIuJ$,
\begin{eqnarray}
\label{eq:ETomeEomeavr}
\ETome&\ge& 
\Eomeavr,
\end{eqnarray}
and 
\begin{eqnarray}
\label{eq:ETomeEomeI}
\ETome 
\ge  \EomeI, \ 
\ETome 
\ge \EomeJ.
\end{eqnarray}
\end{lem}
\proof 
The inequality (\ref{eq:half})
 follows directly from  the definitions.
The optimal decomposition 
 of $\ETome$ is given by 
some 
$\ome = \sum \lami \omeie $
 such that all   $\omeie$ are pure and  even. 
Since each  $\omeie$
  satisfies   (\ref{eq:symmetric}),
(\ref{eq:ETomeEomeavr}) and 
(\ref{eq:ETomeEomeI}) follow.
\proofend
\ \\

The inequalities 
    (\ref{eq:half}) and  (\ref{eq:ETomeEomeI})
 are  exact for $\vpl$ of $\lambda\ne1$  in $\S$ 
\ref{sec:GAUGE},  since 
  it  is always  separable for 
$(\AlI, \AlIprime)$ and  for $(\AlJ, \AlJprime)$
 hence  $\EvplI=\EvplJ=0$,  while 
 for  the case of  (\ref{eq:lam2})
 it is  non-separable  for  $(\AlI, \AlJ)$
and hence  both $\ETvpl$ and $\Evplavr$ should be  nonzero.

The  noneven pure state $\varrho$ with its asymmetric 
 marginal states
(\ref{eq:varrho}) 
 gives  
$\EvarrhoI=0$,
$\EvarrhoJ=\log 2$, and
$\Evarrhoavr=1/2(\log 2)$.
Hence 
\begin{eqnarray*}
\Eomeavr
\ge  \EomeI \ {\text{and}}\ \ 
\Eomeavr
\ge  \EomeJ
\end{eqnarray*}
 is not satisfied in general.
%

We can now generalize Proposition \ref{pro:NAR}
 to the case including  noneven states,  assuming 
 additionally that  
 the systems are  finite dimensional.
\begin{pro}
\label{pro:GNA}
Let $\I$ and $\J$ be finite subsets and
 $\ome$ be a      state 
on  $\AlIuJ$.
If it is  separable 
 for  the CAR pair $\AlI$ and $\AlJ$, then  
 so it is   for the tensor product pair 
 $\AlI$ and    $\AlIprime$.
\end{pro}
\proof
If it is separable, then 
$\Eomeavr=0$.
Hence by  (\ref{eq:half}),
$\EomeI=0$.
This is equivalent to the separability
 of 
 $\ome$ for  ($\AlI$,   $\AlIprime$).
\proofend
\ \\

By Propositions 
\ref{pro:evencrit} and 
\ref{pro:nonevencrit}, 
both 
$\ETome$
and $\Eomeavr$ serve   characterization 
 of separable states  for $(\AlI, \AlJ)$.
We do not know whether the inequality
(\ref{eq:ETomeEomeavr}) can be  strict.
\ \\
\ \\
%

\end{document}